\documentclass[sigconf]{acmart}
\AtBeginDocument{%
  }

\setcopyright{acmlicensed}
\copyrightyear{2025}
\acmYear{2025}
\acmDOI{XXXXXXX.XXXXXXX}
\acmConference[TSMO at KDD '25]{Two-sided Marketplace Optimization: 
  Search, Discovery, Matching, Pricing \& Growth 
  in conjunction with KDD Conference (KDD 2025)}{August 4,
  2025}{Toronto, Canada}
\acmISBN{XXX-X-XXXX-XXXX-X/2025/08}

\usepackage{adjustbox}

\begin{document}

\title{Zero-Shot Retrieval for Scalable Visual Search in a Two-Sided Marketplace}

\author{Andre Rusli}
\email{andre.rusli@mercari.com}
\orcid{0000-0002-7907-2394}
\affiliation{%
  \institution{Mercari, Inc.}
  \city{Tokyo}
  \country{Japan}
}

\author{Shoma Ishimoto}
\email{s-a-ishimoto@mercari.com}
\orcid{0009-0007-4871-2030}
\affiliation{%
  \institution{Mercari, Inc.}
  \city{Tokyo}
  \country{Japan}
}

\author{Sho Akiyama}
\email{s-akiyama@mercari.com}
\orcid{0009-0008-0668-5182}
\affiliation{%
  \institution{Mercari, Inc.}
  \city{Tokyo}
  \country{Japan}
}

\author{Aman Kumar Singh}
\email{itsaman141@mercari.com}
\orcid{0009-0002-6196-9344}
\affiliation{%
  \institution{Mercari, Inc.}
  \city{Tokyo}
  \country{Japan}
}

% \renewcommand{\shortauthors}{Rusli et al.}

\begin{abstract}
Visual search offers an intuitive way for customers to explore diverse product catalogs, particularly in consumer-to-consumer (C2C) marketplaces where listings are often unstructured and visually driven. This paper presents a scalable visual search system deployed in Mercari’s C2C marketplace, where end-users act as buyers and sellers. We evaluate recent vision-language models for zero-shot image retrieval and compare their performance with an existing fine-tuned baseline. The system integrates real-time inference and background indexing workflows, supported by a unified embedding pipeline optimized through dimensionality reduction. Offline evaluation using user interaction logs shows that the multilingual SigLIP model outperforms other models across multiple retrieval metrics, achieving a 13.3\% increase in nDCG@5 over the baseline. A one-week online A/B test in production further confirms real-world impact, with the treatment group showing substantial gains in engagement and conversion, up to a 40.9\% increase in transaction rate via image search. Our findings highlight that recent zero-shot models can serve as a strong and practical baseline for production use, which enables teams to deploy effective visual search systems with minimal overhead, while retaining the flexibility to fine-tune based on future data or domain-specific needs.
\end{abstract}



\received{15 May 2025}
\received[revised]{10 July 2025}
\received[accepted]{27 June 2025}

\maketitle

\section{Introduction}

\begin{figure}[h]
  \centering
  \includegraphics[width=\linewidth]{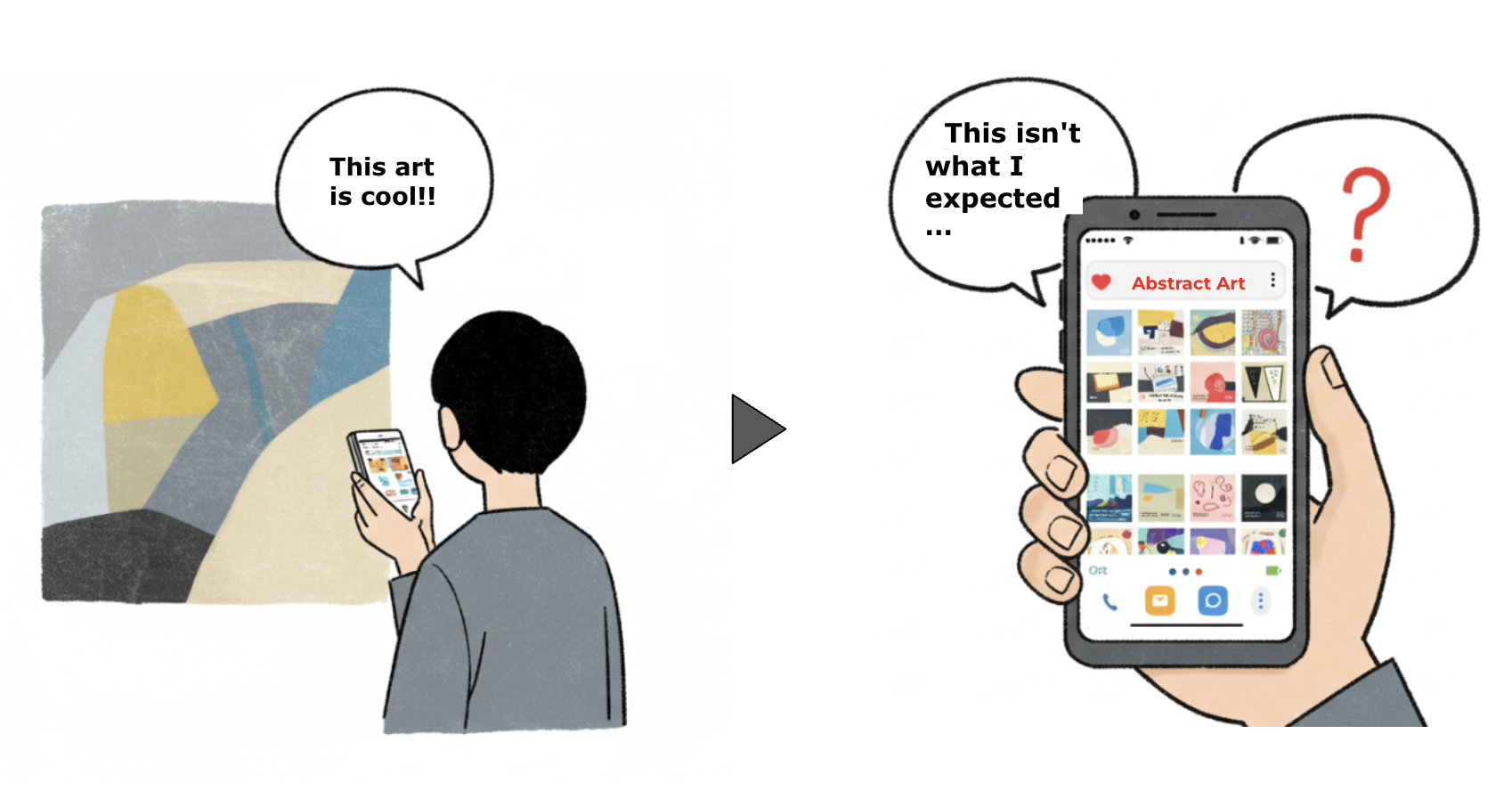}
  \caption{An example use-case where visual search can benefit potential buyers.}
  \Description{Example use-case showing how visual search improves discovery of visually distinctive items.}
  \label{fig:user-needs}
\end{figure}

C2C (consumer-to-consumer) marketplace platforms like Mercari are two-sided ecosystems where end-users can participate as buyers and sellers. Unlike traditional B2C platforms, most sellers on C2C marketplaces are not professional merchants but everyday individuals listing second-hand or surplus items. This leads to a diverse and often inconsistent product catalog, where items of the same type may have varied naming conventions, category assignments, and visual quality. As a result, navigating the catalog—whether to make a purchase or to research similar listings before selling—can be challenging for the customers. This study focuses on Mercari Japan’s C2C marketplace, which serves a large domestic user base with over 20 million monthly active users and covers a broad range of product categories.

While text-based search engines remain the default method for product discovery, they often struggle in such environments. Many items listed on C2C platforms are difficult to describe precisely in words or are identified primarily by visual traits (e.g., fashion, character goods, or collectibles). This is especially problematic when sellers use incomplete or ambiguous descriptions. As illustrated in \autoref{fig:user-needs}, visual search provides an intuitive alternative, helping end-users locate visually similar items without clean or structured textual metadata. It also supports selling workflows by enabling sellers to estimate market value through image-based lookup before listing their items.

Several leading e-commerce companies have adopted visual search solutions tailored to their marketplaces \cite{du2022amazon, shiau2020shop, yang2017visual}, often relying on fine-tuned image embedding models. While effective in controlled settings, these approaches require significant manual effort, labeled data, and infrastructure maintenance—factors that make them costly and potentially brittle when scaled to the messy, ever-evolving landscape of C2C platforms. Despite the growing interest in zero-shot and pre-trained approaches, existing literature lacks a systematic evaluation of whether recent vision-language models can achieve competitive or superior performance in large-scale marketplace settings without fine-tuning.

Recent advances in foundational models—particularly those trained on large-scale image-text pairs, such as CLIP \cite{radford2021clip} and SigLIP \cite{zhai2023sigmoid}—have shown strong generalization across domains. These models offer robustness to distribution shifts and eliminate the need for extensive downstream training, making them attractive for zero-shot retrieval. Prior studies have found pre-trained models effective for various e-commerce tasks such as category classification \cite{shin2022clip}, attribute extraction \cite{yang2017visual}, and category-to-image retrieval \cite{hendriksen2022extending}, and new benchmarks are emerging for region-specific model evaluation \cite{clipeval2025github}.

The main contributions of this work are threefold. First, we present a scalable visual search pipeline designed to address the specific challenges of C2C marketplaces, where end-users act as both buyers and non-professional sellers. Second, we investigate the potential of zero-shot image retrieval using recent vision-language models—specifically SigLIP—and compare their performance to a fine-tuned production baseline. While the multilingual SigLIP model performs favorably in our setting, the primary goal is to assess the practicality of zero-shot approaches in dynamic marketplace environments, where fine-tuned models may be less robust to distributional shifts. Third, we comprehensively evaluate the system, including offline metrics based on historical interaction data, online A/B testing to measure user impact in a production setting, and qualitative analysis to better understand the model’s behavior and limitations. 

\begin{figure}[h]
  \centering
  \includegraphics[width=\linewidth]{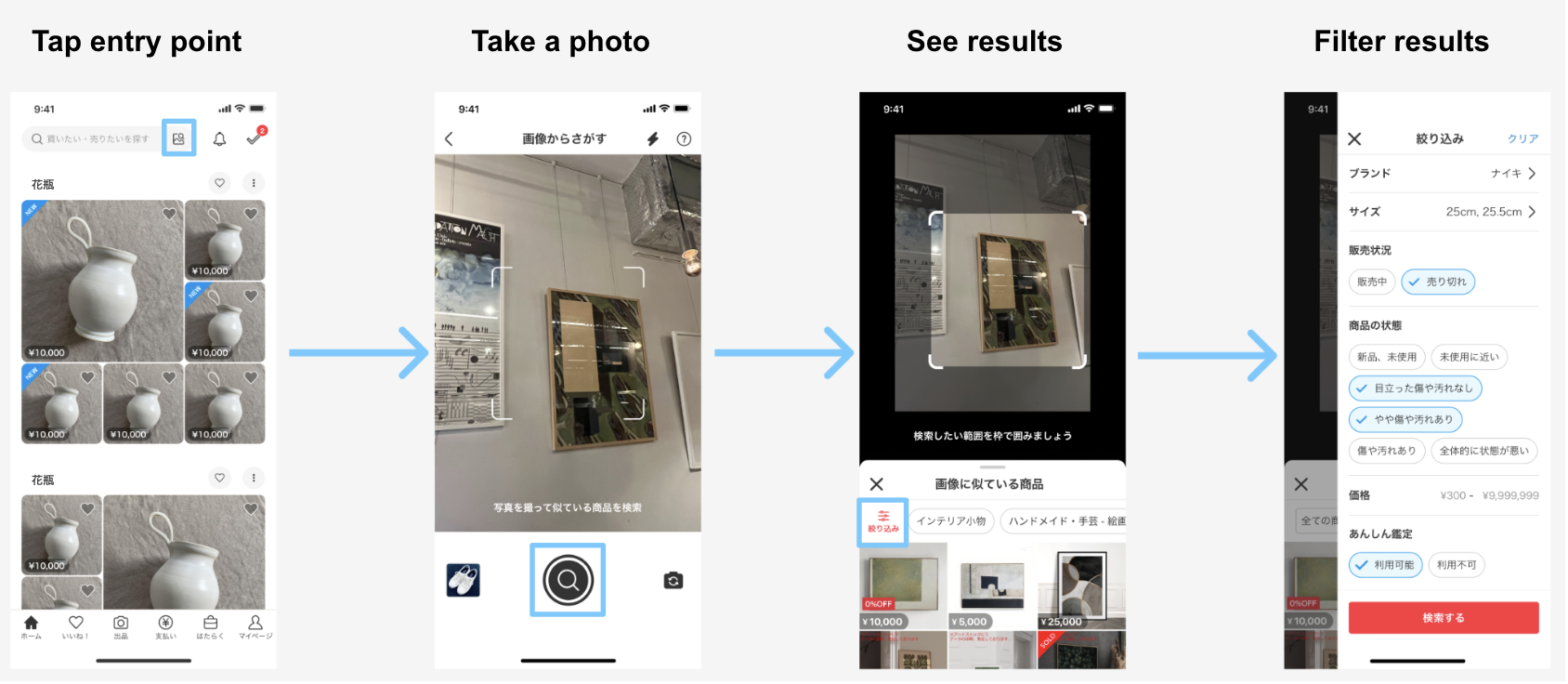}
  \caption{The UX flow of the proposed visual search feature.}
  \Description{A flow diagram illustrating the user journey through the visual search feature on the mobile app.}
  \label{fig:visual-search-ux}
\end{figure}

\section{Methodology}

\autoref{fig:visual-search-ux} illustrates the user experience flow of the visual search system within our mobile platforms. End-users would be able to enter the visual search screen from one of the entry points, upload an image, and the system would return the resulting items. We have implemented a "Similar Looks Component" recommending visually similar listings to customers viewing a listed product in one of our previous works. This feature has successfully helped customers who have viewed a certain listing to discover other listings with comparable visual characteristics. This feature is powered by a vision model fine-tuned on millions of platform-specific images. While re-purposing this model for our new visual search system seems intuitive, the distribution of input images differs substantially. The Similar Looks Component processes curated, seller-optimized listing images as inputs, whereas the proposed visual search system must handle raw, user-uploaded images from diverse sources. Nevertheless, evaluating this legacy fine-tuned model's performance in the new context merits investigation, which we treat as the "baseline" model and explore in subsequent sections.

Our experimental approach began with a limited release of our visual search system's initial version, powered by our legacy fine-tuned model (in-house baseline) already deployed for the Similar Looks Component described above. This strategic roll-out allowed us to collect an initial set of usage logs, gaining valuable insights into customer interaction patterns, user needs, and potential implementation challenges. Following this preliminary deployment, we systematically gathered and analyzed logs and behavioral data to construct a comprehensive offline evaluation dataset. Specifically, we gathered search session logs containing query images uploaded by the users, the list of retrieved item images, and the list of item images the users interacted with, which will then be treated as the true labels in the evaluation set. This enabled the team to conduct rigorous comparison of various models and methodologies in the subsequent phases.

Using the above logs to build an evaluation set, we conducted comparative zero-shot performance analysis using three base models and our pre-existing fine-tuned model:

\begin{itemize}
\item{\textit{baseline}}: A pre-existing vision model, fine-tuned with Mercari's listing images for the existing Similar Looks Component.
\item {\textit{clip-japanese-base}}: A CLIP-based model \cite{clip-japanese-base} specifically adapted for Japanese texts, scoring the highest averaged score in the LLM-jp evaluation benchmark.
\item{\textit{siglip-base-patch16-256-multilingual}}: A language-image pre-trained model \cite{zhai2023sigmoid} with sigmoid loss supporting multiple languages.
\item{\textit{dinov2-large}}: A self-supervised vision model \cite{oquab2023dinov2} known for robust representations.
\end{itemize}

These models are selected considering factors such as size, which directly impacts production latency and operational costs, and documented performance in recent literature and benchmarks. We evaluated each model's performance using standard information retrieval metrics, including nDCG@k, precision@k, and recall@k. The best-performing model is chosen for an online A/B test to validate its effectiveness in the production environment. We then conduct a comprehensive analysis of key performance indicators between test variants, supplemented by detailed manual qualitative assessments to develop a nuanced understanding of model behavior differences. This multi-stage evaluation process culminated in selecting the optimal variant for full-scale production deployment. The subsequent section presents detailed results and discusses our findings.

\section{Results and Discussion}

\subsection{Zero-shot Retrieval Performance}

\subsubsection{Quantitative Experiments}
We constructed an evaluation set from event logs collected during our initial limited release phase. Each log entry comprised session ID, query image URL, retrieved item ID, corresponding item image URL, and a binary tap interaction indicator as our relevance proxy. User tap behavior provides an implicit feedback signal that is widely accepted as a relevance indicator in recommendation evaluation. We experimented with various number of samples to get consistent results. 

\autoref{tab:quantitative_performance_comparison} compares the performance of the four evaluated models on 10,000 randomly sampled search sessions, containing 193,072 unique images. The multilingual SigLIP model (\textit{siglip-base-patch16-256-multilingual}) outperformed all alternatives with an nDCG@5 of 0.578—a 13.3\% improvement over our in-house baseline model (\textit{baseline})—and achieved superior precision and recall across all retrieval metrics. DINOv2 ranked second, rivaling the baseline model, while the Japanese CLIP model (\textit{clip-japanese-base}) trailed despite its language specialization.

\begin{table}
 \caption{Comparison of results of different models on the offline evaluation dataset}
 \label{tab:quantitative_performance_comparison}
 \centering
 \adjustbox{max width=\columnwidth}{
 \begin{tabular}{lccccccc}
   \toprule
   Model & nDCG@5 & P@5 & P@3 & P@1 & R@5 & R@3 & R@1 \\
   \midrule
   \textit{baseline} & 0.510 & 0.190 & 0.235 & 0.332 & 0.551 & 0.432 & 0.228 \\
   \textit{clip-japanese-base} & 0.472 & 0.175 & 0.218 & 0.305 & 0.516 & 0.398 & 0.213 \\
   \textit{siglip-base-patch16-256-multilingual} & \textbf{0.578} & \textbf{0.212} & \textbf{0.267} & \textbf{0.386} & \textbf{0.634} & \textbf{0.499} & \textbf{0.271} \\
   \textit{dinov2-large} & 0.526 & 0.190 & 0.243 & 0.354 & 0.565 & 0.457 & 0.246 \\
   \bottomrule
 \end{tabular}
 }
\end{table}

We also analyzed computational efficiency by measuring the average inference time per image across all models on an A100 GPU. The baseline model, multilingual SigLIP, and Japanese CLIP exhibited comparable performance, with average inference times of approximately 0.033 seconds per image. In contrast, DINOv2 Large was substantially slower at around 0.048 seconds per image, about 47\% longer than the other models, which we attribute to its larger embedding dimension.

The empirical evidence indicates that multilingual SigLIP provides the optimal balance between retrieval effectiveness and computational efficiency for our visual search application. Its superior performance across all relevance metrics without additional computational overhead makes it a promising candidate for production deployment.

\subsubsection{Qualitative Assessments}

To complement the quantitative evaluations, we conducted a qualitative assessment comparing the highest-scoring model, the multilingual SigLIP, with our baseline model using authentic user-uploaded images. Our analysis revealed that the multilingual SigLIP consistently produced more semantically relevant and contextually accurate retrievals across diverse image categories.

Its robust recognition capabilities were particularly evident in cases requiring fine-grained understanding of visual features, such as identifying specific characters in user-uploaded images, as shown in \autoref{tab:qualitative_performance_comparison_good}. Unlike the in-house baseline model, which often struggled to disambiguate similar-looking objects and returned only a handful of accurate results, it could retrieve highly relevant items consistently. This indicates a stronger ability to generalize to nuanced visual contexts. Additionally, the multilingual SigLIP demonstrated resilience to image noise, outperforming the baseline model in scenarios where nonsemantic features, such as font styles or text, might otherwise mislead the retrieval process. For example, the baseline model occasionally prioritized superficial visual similarities, yielding semantically unrelated and visually adjacent results.

\begin{table}
 \caption{Comparison of actual images retrieved by the two models}
 \label{tab:qualitative_performance_comparison_good}
 \centering
 \adjustbox{max width=\columnwidth}{
 \begin{tabular}{lccccccc}
   \toprule
   Item Category & Query Image & Baseline Model & Multilingual SigLIP  \\
   \midrule
   Character Goods & 
   \includegraphics[width=0.3\linewidth]{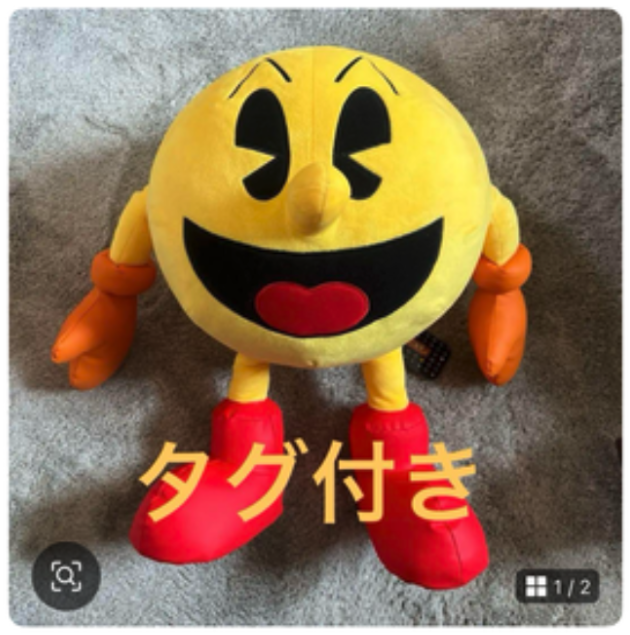} & 
   \includegraphics[width=0.3\linewidth]{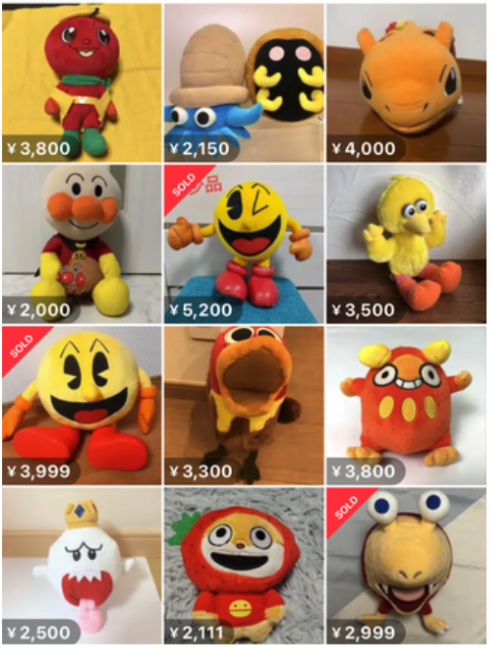} & 
   \includegraphics[width=0.3\linewidth]{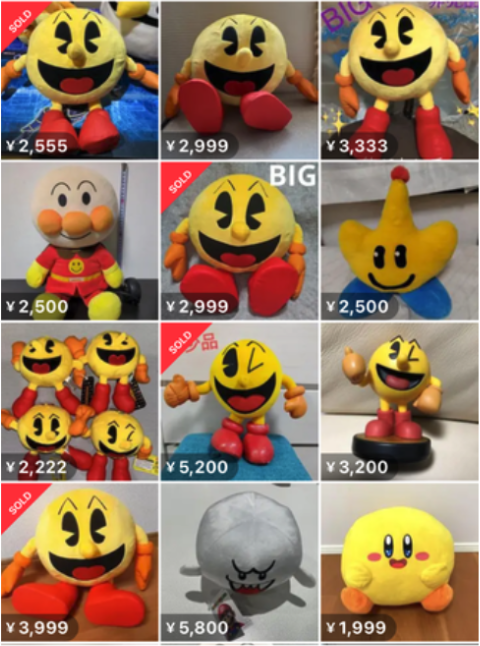} \\
   Drinks & 
   \includegraphics[width=0.3\linewidth]{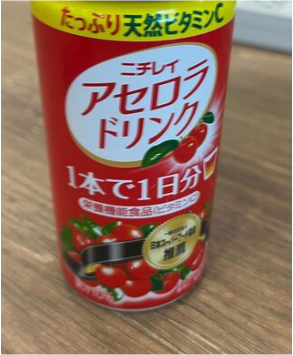} & 
   \includegraphics[width=0.3\linewidth]{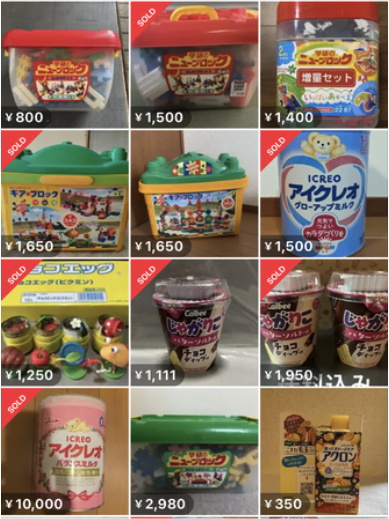} & 
   \includegraphics[width=0.3\linewidth]{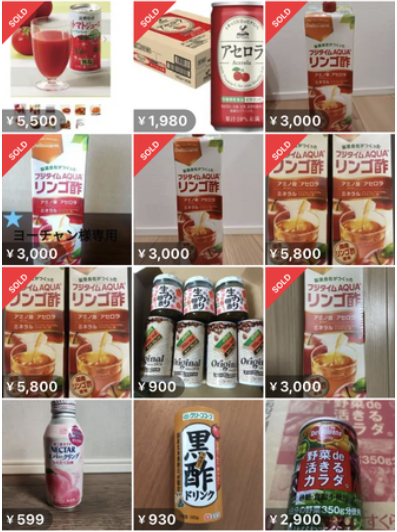} \\
   Sneakers & 
   \includegraphics[width=0.2\linewidth]{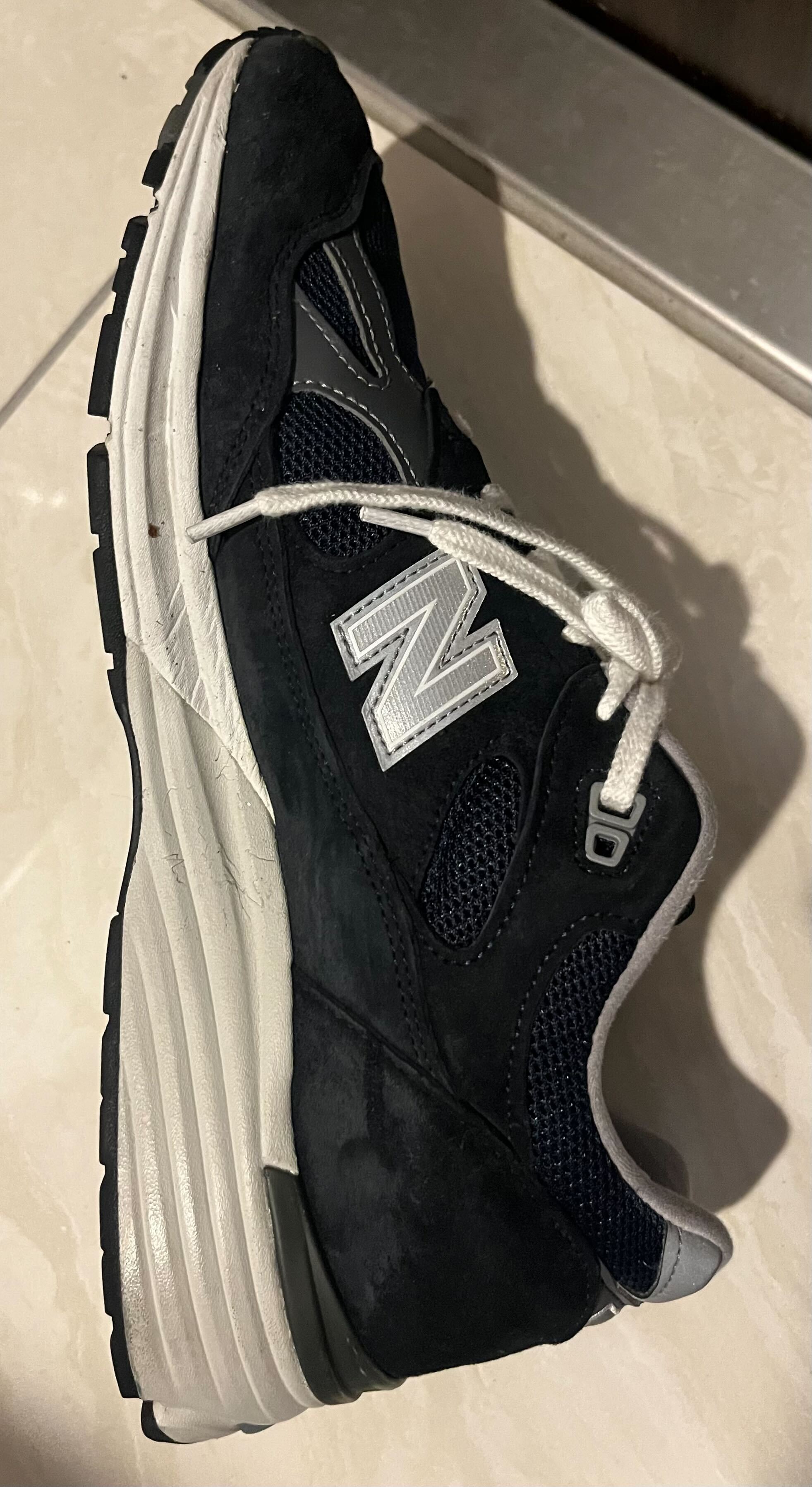} & 
   \includegraphics[width=0.3\linewidth]{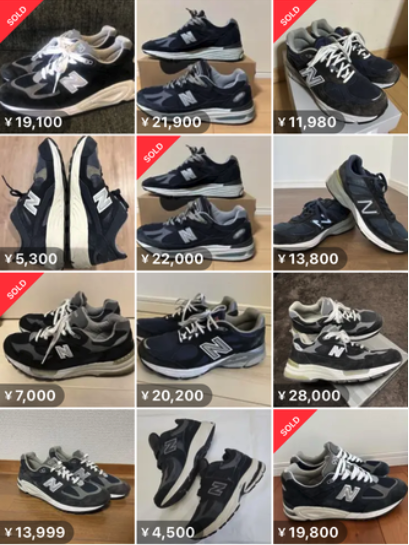} & 
   \includegraphics[width=0.3\linewidth]{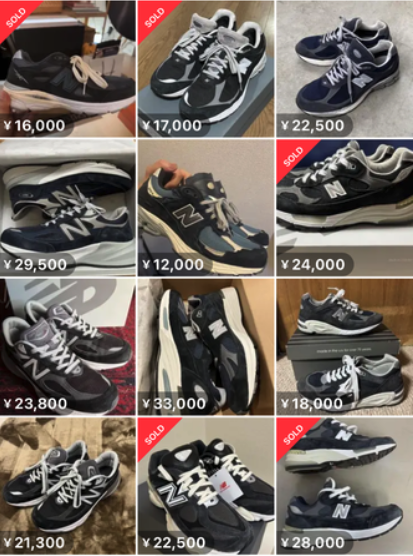} \\
   \bottomrule
 \end{tabular}
 }
\end{table}

These observations underscore the multilingual SigLIP model's superior zero-shot performance and reliability in addressing real-world visual search challenges. It achieves higher recognition accuracy, improved semantic relevance, and robust generalization across a variety of user queries, demonstrating potential to provide a more intuitive and effective image search experience in production settings.

\subsection{System Architecture}

\begin{figure}[h]
  \centering
  \includegraphics[width=\linewidth]{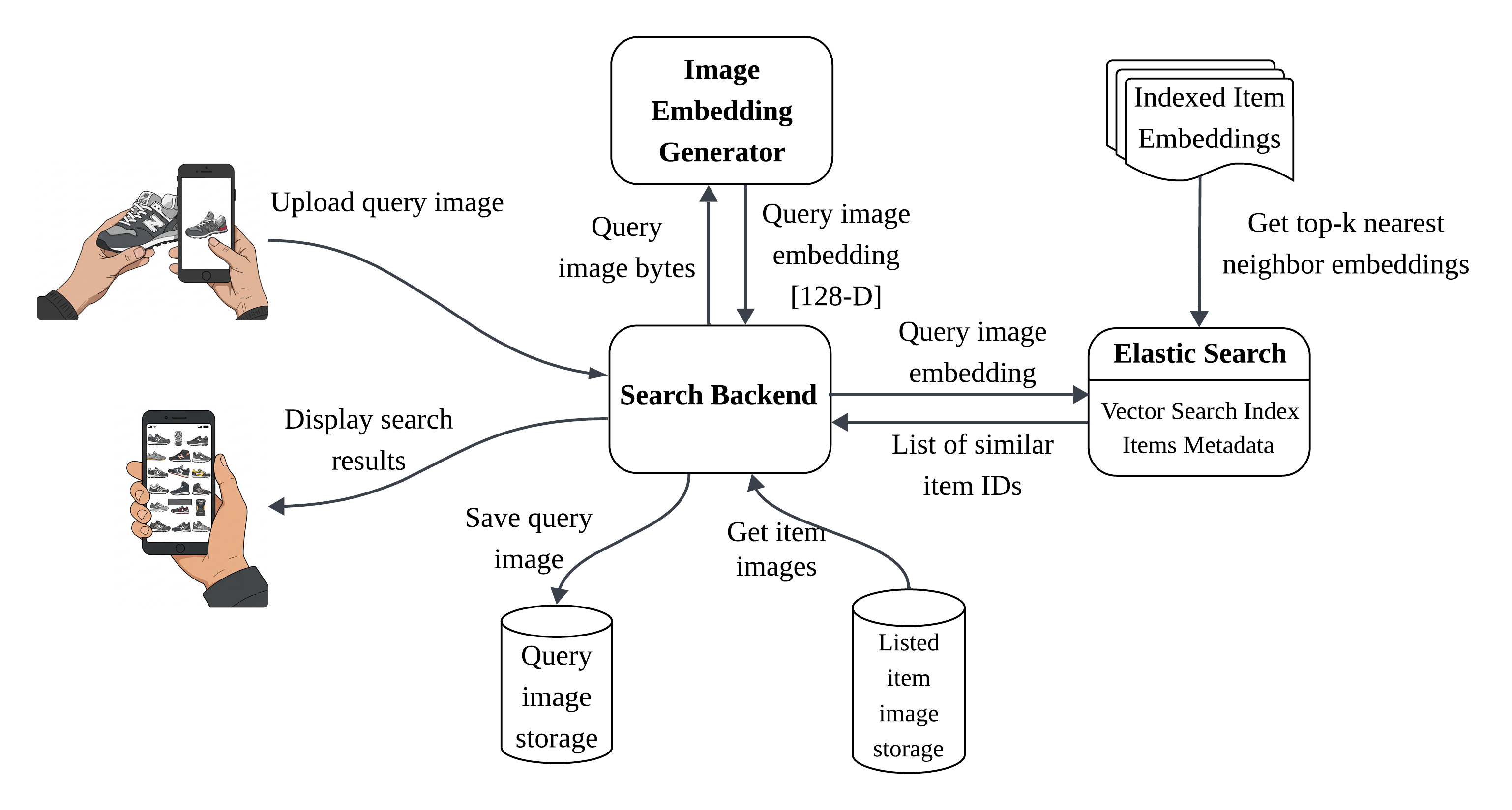}
  \caption{System flow of when a user upload a search query image}
  \Description{A diagram illustrating the how the system works when a user uploads an image as a search query.}
  \label{fig:system-flow-user-search}
\end{figure}

We design the proposed visual search system to support both real-time image-based retrieval and continuous background catalog indexing in a scalable and efficient manner. This allows us to efficiently handle both indexing processes in the background and real-time inference when a user uploads an image as a search query. As illustrated in \autoref{fig:system-flow-user-search}, when a user uploads a search query image, the image is first received by the search backend, which forwards the raw image bytes to the image embedding generator. The generator produces a compact 128-dimensional embedding that captures the visual semantics of the input image. This embedding is then sent to the Elastic Search service, where it is compared against a previously indexed set of embeddings using approximate nearest neighbor search. The top-k most similar item IDs are returned to the backend, which retrieves the corresponding item images and metadata from storage. Finally, the search results are displayed to the user, and the query image is then stored for future analysis and continuous evaluation. 

In parallel with the user-facing query processing, the system maintains an up-to-date index of the item catalog through two indexing workflows to ensure that all new listings are quickly indexed and made available for visual search. This also enables efficient batch processing of the entire catalog when model updates occur. Separating concerns between microservices, dataflow components, and the inference server provides flexibility for independent scaling and updates to different system parts.

\begin{itemize}
\item {Offline Indexing}: This batch pipeline is executed periodically to process and pre-index many item listings created within a specified time window. The initial indexing mechanism is to bootstrap the visual search system, ensuring that a substantial portion of the item catalog is searchable from the outset.
\item {Online Indexing}: The system also includes an online indexing pipeline to support real-time searchability of newly listed or updated items. When a seller lists a new item or modifies an existing one, a microservice triggers and processes an event, which saves the associated image to a central image database. 
\end{itemize}

\begin{figure}[h]
  \centering
  \includegraphics[width=\linewidth]{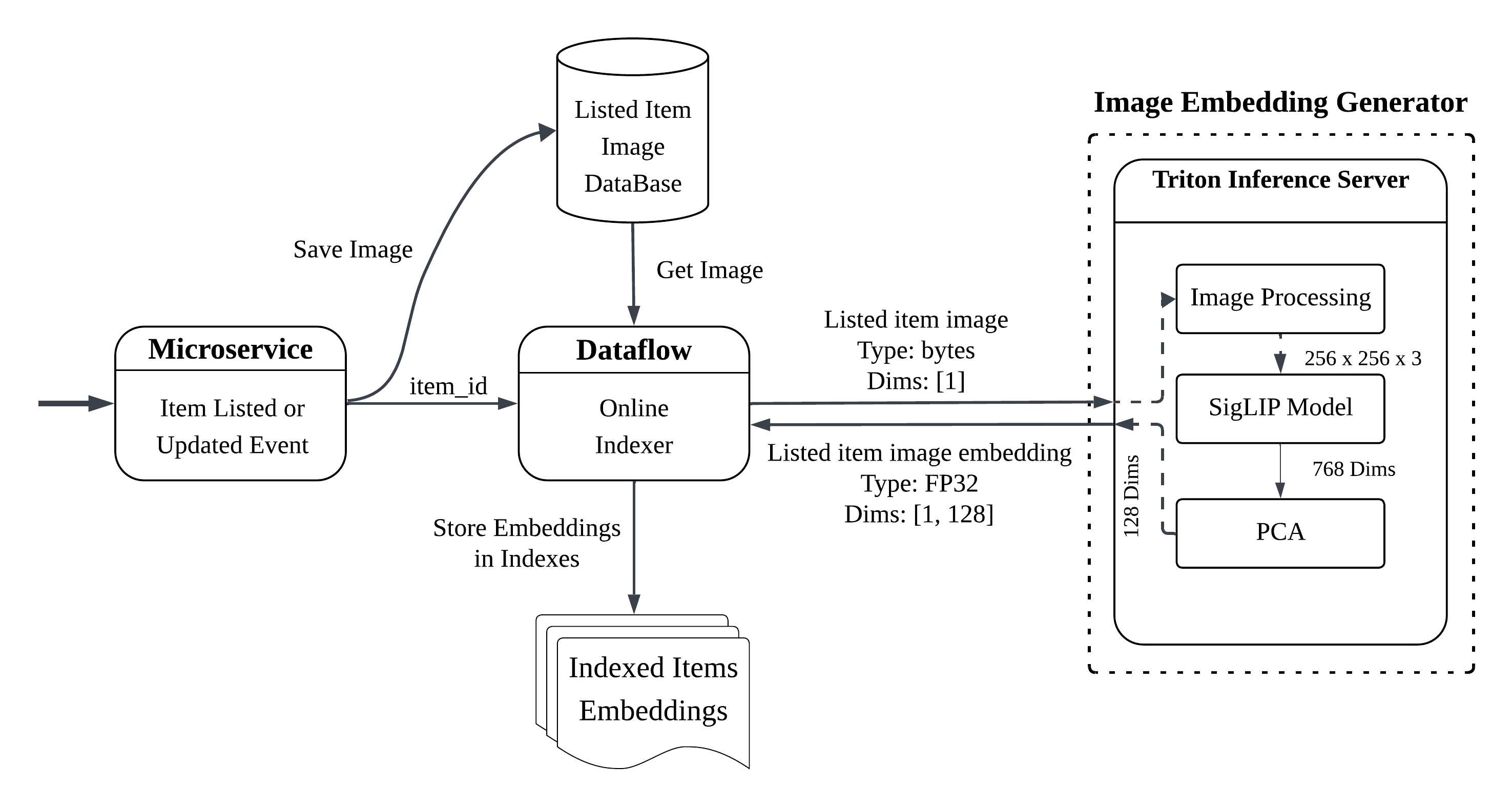}
  \caption{System flow of when listings are created or updated by sellers}
  \Description{A diagram illustrating how the system works when new listings are newly created or updated by sellers.}
  \label{fig:system-flow-online-indexer}
\end{figure}

\autoref{fig:system-flow-online-indexer} describes the online indexing process in more detail. When a seller lists a new item or updates an existing one, a dedicated microservice emits and captures an event, extracts the \texttt{item\_id}, and stores the associated image in a centralized image database. The online indexer then retrieves this image and forwards it to the Image Embedding Generator, which is hosted on a Triton Inference Server.

Within the Image Embedding Generator, the image undergoes a standardized embedding pipeline consisting of the following steps:

\begin{enumerate}
  \item {Image Preprocessing}: The raw image is resized and normalized to a uniform resolution of $256 \times 256 \times 3$ to match the input format expected by the embedding model.
  \item {Feature Extraction}: The preprocessed image is passed through the multilingual SigLIP model, which produces a 768-dimensional embedding vector representing the semantic features of the image.
  \item {Dimensionality Reduction}: The high-dimensional embedding is projected into a 128-dimensional space using Principal Component Analysis (PCA), balancing semantic fidelity with computational and storage efficiency.
\end{enumerate}

The resulting 128-dimensional vector (in FP32 format) is returned to the online indexer and stored in the vector index alongside existing item embeddings. This enables immediate retrieval in response to user image queries and ensures that the visual search index remains up-to-date with minimal latency. 

The combination of real-time query processing and robust indexing workflows enables the system to deliver up-to-date and responsive visual search experience at scale. To validate the practical effectiveness of the proposed architecture, particularly the integration of the multilingual SigLIP model, we conducted an online A/B test to assess its impact in a live production environment, which is described in the next section below.

\subsection{Online A/B Test and Practical Implications}

Following the promising results from the offline evaluation, we conducted a one-week online A/B test to assess the impact of the multilingual SigLIP model in a production setting. Users are randomly assigned to either the control group (50\%) with the baseline model or the treatment group (50\%) with the multilingual SigLIP model. The experiment aimed to measure improvements across key image search performance metrics, some shown in \autoref{tab:ab_test_results}.

\begin{table}
  \caption{Online A/B test results between the baseline model (Control) and the multilingual SigLIP model (Treatment)}
  \label{tab:ab_test_results}
  \centering
  \adjustbox{max width=\columnwidth}{
    \begin{tabular}{lccc}
      \toprule
      \textbf{Metric} & \textbf{Control} & \textbf{Treatment} & \textbf{Treatment / Control} \\
      \midrule
      Average Transaction per User via Image Search & 0.00308 & 0.00435 & +40.9\% \\
      Buy Conversion Rate via Image Search          & 0.00256 & 0.00344 & +34.1\% \\
      Item View Count per User via Image Search     & 3.16    & 4.64    & +46.6\% \\
      \bottomrule
    \end{tabular}
  }
\end{table}

The results showed that the multilingual SigLIP model significantly boosted user engagement and purchase behavior. The treatment group exhibited a 40\% increase in Average Transactions per User via Image Search, indicating higher purchase rates driven by image-based queries. Additionally, a 34\% increase in Buyer Conversion Rate via Image Search and a 46.6\% increase in Item View Count per User via Image Search were observed, underscoring greater user engagement and conversion efficacy.

The following month after the above A/B test concluded, the image search is utilized by approximately 1.5 million users per month (around 7\% of monthly active users in Mercari) and contributes to increased purchases by creating new matching experiences. It is used in categories such as fashion, talent, and character goods, meeting the need to search for products based on appearances. It is also used for listing purposes, where sellers take photos of their belongings before listing them and use them for market price research to determine how much they can sell for, thereby contributing to an increase in the number of listings.

However, there are also some cases where both models suffer in performance. One prominent example is when it needs to recognize a specific person based on the facial features. This needs to be improved since, based on in-depth interviews with customers, it is well-known that, especially in anime, character goods, and entertainment (ACGE) categories, the ability to retrieve precisely the same human or animated characters is crucial to the customers. \autoref{tab:qualitative_performance_comparison_bad} shows an example of such cases, where images with similar features are retrieved, but not always of the same person. Some features derived from the images are recognized correctly, such as the angle of the pose and the hand gestures. However, it shows completely different idol members even though we found that there are other items with the same person’s image that the model failed to retrieve, which may hurt the user experience using our image search system since many customers expect that it shows the same person, even with a different pose. 

\begin{table}
 \caption{Examples where the model struggles to retrieve the relevant person images}
 \label{tab:qualitative_performance_comparison_bad}
 \centering
 \adjustbox{max width=\columnwidth}{
 \begin{tabular}{lccccccc}
   \toprule
   Query Image & Top-6 Similar Images \\
   \midrule
   \includegraphics[width=0.2\linewidth]{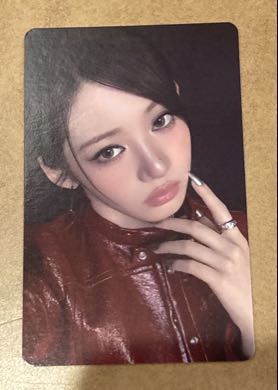} & \includegraphics[width=1.2\linewidth]{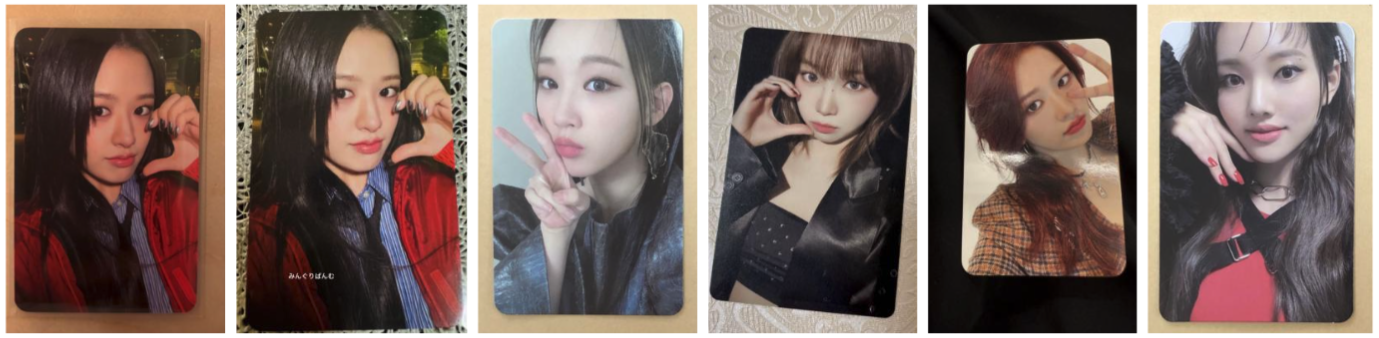} \\
   \includegraphics[width=0.2\linewidth]{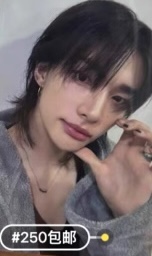} & \includegraphics[width=1.2\linewidth]{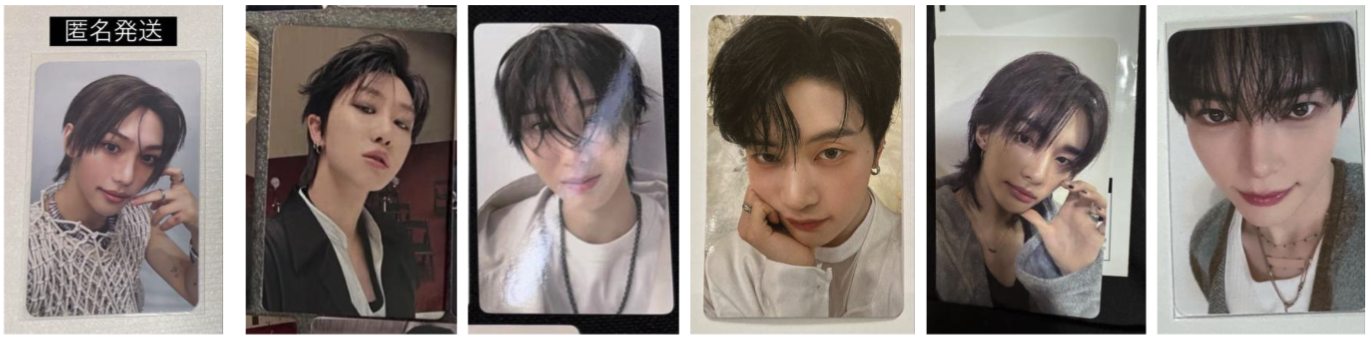} \\
   \bottomrule
 \end{tabular}
 }
\end{table}

Lastly, we experimented with various dimensionality reduction strategies using Principal Component Analysis (PCA) to optimize the embedding size for production use. Through empirical evaluation, we found that reducing the original 768-dimensional SigLIP embeddings to 128 dimensions provided the best trade-off between retrieval performance and system efficiency. This configuration preserved sufficient semantic information for accurate image retrieval while yielding significant infrastructure benefits: approximately 35–40\% reduction in Elasticsearch query latency and around 83\% decrease in memory usage. These improvements contributed to faster response times and lower operational costs, enabling scalable deployment without compromising search quality. For a detailed summary of the observed efficiency gains, see \autoref{tab:dimensionality_reduction_impact}.

\begin{table}
  \caption{Impact of dimensionality reduction}
  \label{tab:dimensionality_reduction_impact}
  \centering
  \adjustbox{max width=\columnwidth}{
    \begin{tabular}{lccc}
      \toprule
      \textbf{Metric} & \textbf{768-D Embeddings} & \textbf{128-D Embeddings} & \textbf{Relative Change} \\
      \midrule
      p50 Query Latency & 100 ms & 60 ms & $\sim$40\% reduction \\
      Memory per Vector (float32) & 3.0 KB & 0.5 KB & $\sim$83\% reduction \\
      Index Size per 1M Vectors & 3.0 GB & 0.5 GB & $\sim$83\% reduction \\
      \bottomrule
    \end{tabular}
  }
\end{table}

\section{Conclusion and Future Work}

This paper presents a scalable visual search system for a large-scale C2C marketplace. It leverages zero-shot retrieval using multilingual SigLIP embeddings and dimensionality reduction to balance effectiveness and efficiency. Through extensive offline evaluation, online A/B testing, and qualitative analysis, we demonstrate that the proposed system improves semantic retrieval quality and user engagement while significantly reducing infrastructure costs. The deployment of this system has led to increased transaction volume, higher conversion rates, and broader adoption of image-based interactions among customers in Mercari.

Despite the promising results, several challenges remain. The system notably struggles with precise identity matching, particularly in categories such as entertainment and character goods, where customers expect not only visually similar, but exact product matches. Addressing this will require further specialization, such as incorporating identity-aware or region-specific fine-tuning, or integrating multimodal signals including text and user feedback. Nonetheless, the findings from this work highlight the practicality of zero-shot vision-language models for real-world retrieval tasks and suggest that scalable, general-purpose embeddings can serve as a strong foundation for continued innovation in visual search systems.

Future work will focus on several directions, including: improving finer-grained retrieval through multimodal embeddings, model adaptation, and re-ranking mechanisms; exploring personalization strategies that leverage user history and intent to refine search results; and evaluating longer-term user behavior changes and business impact beyond short-term A/B tests to understand better retention and listing behavior driven by visual search.

\begin{acks}
We would like to express our sincere gratitude to the engineering members of the Search, AI/LLM, and ML Platform teams for their outstanding contributions throughout the development of this work. Their deep technical expertise, collaborative spirit, and relentless focus on scalability and performance were instrumental in building and deploying the system. We also extend our heartfelt thanks to our cross-functional partners, including designers, product managers, and UX researchers, whose thoughtful input ensured a seamless and user-centric experience for our customers. Special appreciation goes to the Business Intelligence (BI) team for their analytical support in shaping our evaluation strategy, to the QA team for their rigorous testing and quality assurance, and to the Marketing team for their efforts in driving awareness and adoption. Their dedication and expertise have collectively played a vital role in bringing this work to life.
\end{acks}

\bibliographystyle{ACM-Reference-Format}
\bibliography{mercari-references}

\end{document}